\documentclass[sma4]{snow2e}
\newenvironment{comment}[1]{}{}
\def\beq{\begin{equation}}
\def\eeq{\end{equation}}
\def\bea{\begin{eqnarray}}
\def\eea{\end{eqnarray}}
\def\barr{\begin{array}}
\def\earr{\end{array}}
\newcommand{\sm}{standard model}
\newcommand{\cm}{center of mass}
\newcommand{\xs}{cross section}

\newcommand{\pe}{\mbox{$e^+e^-$}}
\newcommand{\ee}{\mbox{$e^-e^-$}}
\newcommand{\ep}{\mbox{$e^-\gamma$}}
\newcommand{\pp}{\mbox{$\gamma\gamma$}}
\newcommand{\sw}{\mbox{$\sin\theta_w$}}
\newcommand{\cw}{\mbox{$\cos\theta_w$}}
\newcommand{\swt}{\mbox{$\sin^2\theta_w$}}

\begin{document}

\title{\boldmath 
Systematics Effects in $Z'$ Searches at the NLC
}
\author{
Frank Cuypers\thanks{\tt cuypers@pss058.psi.ch} 
\\
{\it Paul Scherrer Institute, CH--5232 Villigen PSI, Switzerland} 
}

\maketitle

%% Get rid of page numbering
\thispagestyle{empty}\pagestyle{empty}

\begin{abstract} 
We investigate the incidence of 
several systematic effects
on the $Z'$ discovery limits
of the NLC.
These include
the initial state radiation
and the systematic errors 
due to 
the imperfect polarization measurement,
the finite detector angular resolution and
the uncertainty on the integrated luminosity.
We focus on three reactions 
involving leptonic couplings:
muon pair production,
Bhabha scattering
and M\o ller scattering.
\end{abstract}

\section{Introduction}

An important feature of the NLC
is the high degree of polarization
which can be obtained for the electron beams.
Beam polarizations exceeding 80\%\
are by now routinely obtained at SLAC
and are steadily improving. 
At the NLC a 90\%\ electron polarization seems a 
quite sensible assumption.
For the positron beam, 
although there are reasonable hopes 
that some practicable technology may be available
by the time the NLC is operating,
at present no scheme for polarizing  positrons
has  been proven to be implementable.

Another fascinating feature of the NLC,
is the possibility to run it in its four different modes:
\pe, \ep, \pp\ and \ee~\cite{e-e-}.
In particular the latter collider mode 
has been shown to have a very high
resolving power in searches for a heavy $Z'$~\cite{ccl}
and studies of its couplings to leptons~\cite{zpee}.
This is mainly due 
to the huge events rates of M\o ller scattering
and the possibility of polarizing both beams.
Similar studies have also been performed 
for the more ``bread-and-butter''
muon pair-production reaction 
of the \pe\ mode of the NLC~\cite{zppe}.
To the best of our knowledge,
the prospects of Bhabha scattering
have never been considered in the context of $Z'$ searches.

\begin{figure}
\hskip-2em
\input{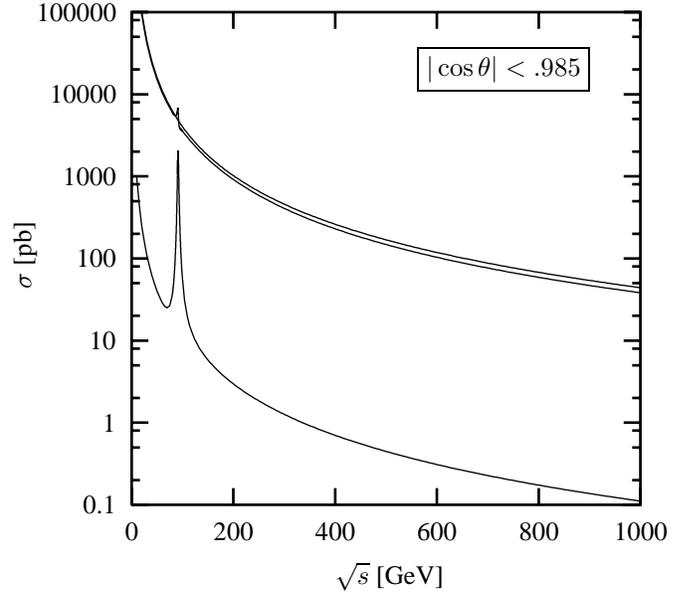}
\caption{
Unpolarized \sm\ \xs s
of the three processes (\ref{mm}--\ref{ee}) 
considered in the text. 
It is easy to disentangle 
which curve corresponds to which reaction!
}
\label{fxs}
\end{figure}

We consider here the following three reactions
involving the exchange of a $Z'$:
\bea
\label{mm} \pe & \to & \mu^+\mu^-
\\
\label{pe} \pe & \to & \pe
\\
\label{ee} \ee & \to & \ee
~.
\eea
In the absence of a $Z'$,
their unpolarized \xs s are plotted as a function of the \cm\ energy
in Fig.~\ref{fxs}.
These curves are obtained
for a polar acceptance angle of {\em ca} $10^o$.
The Bhabha and M\o ller \xs s
exceed those of muon pair-production 
by more than two orders of magnitude
and are thus well suited for precision masurements~\cite{us}.
As we will demonstrate at hand of this $Z'$ analysis,
they are also well adapted to searches for new physics.

There is of course a peak in the \pe\ reactions
if the \cm\ energy is sufficient to produce the $Z'$ on mass shell.
Obviously,
close to this pole
the \ee\ mode cannot compete with the \pe\ mode.
But if the $Z'$ mass exceeds the NLC \cm\ energy
by as little as 20\%,
the discovery potential of the \ee\ mode 
becomes competitive or even superior~\cite{zpee}.
We focus here on the situation
where the $Z'$ is far off-shell.

For each of the reactions (\ref{mm}--\ref{ee})
we analyze here the discovery prospects
of a heavy $Z'$
and study the the influence of four different sytematic effects:
\begin{Itemize}
\item the polarization error
\item the angular resolution
\item the initial state radiation
\item the luminosity error.
\end{Itemize}

\section{Lagrangian}

There are many extensions of the \sm\ 
which predict the existence of extra neutral vector bosons $Z'$.
While most searches are performed within the framework 
of a particular model~\cite{a},
it was advocated
that it is important to perform a model-independent analysis~\cite{l}.
Otherwise we may miss some unexpected kind of new physics
lurking beyond the \sm.
We also adopt this point of view here.

Assuming lepton universality,
the generic lagrangian describing the interaction
of a heavy neutral vector boson $Z'$ 
with the charged leptons $\ell=e,\mu,\tau$
can be written
\beq
L = e~\bar\psi_\ell \gamma^\mu \left( {v_{Z'}} + {a_{Z'}}\gamma_5 \right) \psi_\ell ~{Z'_\mu}
\label{lag}~,
\eeq
where $v_{Z'}$ and $a_{Z'}$ are the vector and axial couplings
normalized to the charge of the electron $e$.
This interaction
mediates both $e^+e^-$ annihilation
and $e^-e^-$ scattering.
%The results of LEP1 constrain any $Z^0$-$Z'$ mixing
%to such an extent \cite{ll}
%that it can safely be ignored here.

Three unknown parameters are involved here:
the $Z'$ mass $m_{Z'}$
and its vector $v_{Z'}$ and axial $a_{Z'}$ couplings.
If the $Z'$ is heavy compared to the collider energy,
the couplings and the mass 
are correlated in such a way that it is difficult to disentangle
a large mass from small couplings
and {\em vice versa}.
Asymptotically,
for $m_{Z'}^2 \gg s$,
the following scaling law applies
\beq
v_{Z'},a_{Z'} \propto m_{Z'} \left( s {\cal L} \right)^{1\over4}
~.
\label{scale}
\eeq
Having determined bounds on the couplings
for a given energy, luminosity and $Z'$ mass,
it is trivial to use this scaling law
to determine the corresponding bounds on the couplings
for different energies, luminosities and $Z'$ masses.

In the limit we consider here
of a heavy $Z'$,
there are thus only two independent parameters
we need to deal with.
We therefore express the discovery limits 
to be set by the three reactions (\ref{mm}-\ref{ee})
in terms of 95\%\ confidence 
exclusion contours 
in the $(v_{Z'},a_{Z'})$ plane,
for a fixed $Z'$ mass.

\section{Cross Sections}

For completeness,
we give here the differential \xs s 
of the three processes (\ref{mm}--\ref{ee})
under consideration.
As we consider the situation where all intermediate bosons
are far off-shell
$m_{Z^0}^2 \ll s \ll m_{Z'}^2$,
we may neglect the width 
of the $Z^0$ and $Z'$.
In terms of 
the fine structure constant $\alpha$
and the Mandelstam variables
$s,\ t,\ u$
we have
\bea
\label{mmdxs}
&&
{d\sigma(\pe \to \mu^+\mu^-) \over dt}
=
{4\pi\alpha^2 \over s^2}\times
\\\nonumber&&\hskip-2em
\left\{
  [RR]
  \left[
    \left( \sum_i R_i^2 {u \over s-m_i^2}
    \right)^2
    + \left( \sum_i L_iR_i {t \over s-m_i^2} \right)^2
  \right]
\right.
\\\nonumber&&\hskip-2em
+~
  [LL]
  \left[
    \left( \sum_i L_i^2 {u \over s-m_i^2}
    \right)^2
    + \left( \sum_i L_iR_i {t \over s-m_i^2} \right)^2
  \right]
~,
\eea

\bea
\label{pedxs}
&&
{d\sigma(\pe \to \pe) \over dt}
=
{4\pi\alpha^2 \over s^2}\times
\\\nonumber&&\hskip-2em
\left\{
  [RR]
  \left[
    \left( \sum_i R_i^2 \left( {u \over s-m_i^2} + {u \over t-m_i^2} \right)
    \right)^2
  \right.
\right.
\\\nonumber&&\hskip8em
\left.
  \left.
    + \left( \sum_i L_iR_i {t \over s-m_i^2} \right)^2
  \right]
\right.
\\\nonumber&&\hskip-2em
+~
  [LL]
  \left[
    \left( \sum_i L_i^2 \left( {u \over s-m_i^2} + {u \over t-m_i^2} \right)
    \right)^2
  \right.
\\\nonumber&&\hskip8em
  \left.
    + \left( \sum_i L_iR_i {t \over s-m_i^2} \right)^2
  \right]
\\\nonumber&&\hskip-2em
\left.
+~
  [LR]
  \left( \sum_i L_iR_i {s \over t-m_i^2} \right)^2
\right\}
~,
\eea

\bea
\label{eedxs}
&&
{d\sigma(\ee \to \ee) \over dt} 
=
{2\pi\alpha^2 \over s^2}\times
\\\nonumber&&\hskip-2em
\left\{
  [RR]
  \left( \sum_i R_i^2 \left( {s \over t-m_i^2} + {s \over u-m_i^2} \right)
        \right)^2
\right.
\\\nonumber&&\hskip-2em
+~
  [LL]
    \left( \sum_i L_i^2 \left( {s \over t-m_i^2} + {s \over u-m_i^2} \right)
         \right)^2
\\\nonumber&&\hskip-2em
\left.
+~
  [LR]
  \left[
    \left( \sum_i L_iR_i {t \over u-m_i^2} \right)^2
  + \left( \sum_i L_iR_i {u \over t-m_i^2} \right)^2
  \right]
\right\}
~.
\eea
The summations are performed over 
the three intermediate bosons
$i=\gamma,Z^0,Z'$.
Their chiral couplings are given by
\bea
R_\gamma = 1
&&
L_\gamma = 1
\\
R_{Z^0} = -{\sw \over \cw}
&&
L_{Z^0} = {1-2\,\swt \over 2\,\sw\cw}
\\
R_{Z'} = v_{Z'} + a_{Z'}
&&
L_{Z'} = v_{Z'} - a_{Z'}
~.
\eea
The polarization coefficients are defined
in terms of the average beam polarizations $P_1$ and $P_2$
by
\bea
\label{rr}
[RR] & = & {1 + P_1 + P_2 + P_1P_2 \over 4}
\\\nonumber\\
\label{ll}
[LL] & = & {1 - P_1 - P_2 + P_1P_2 \over 4}
\\\nonumber\\
\label{lr}
[LR] & = & {1 - P_1P_2 \over 2}
~,
\eea
where $P_1$ and $P_2$
run between -1 and 1
for left to right polarized beams.

\section{Method}

As the Bhabha (\ref{pe}) and M\o ller (\ref{ee}) \xs s
are large,
the systematic errors can easily become far dominant
and spoil the advantage
provided by the large statistics.
The most dangerous systematic error
stems from the luminosity measurement.
However,
the dependence on the luminosity 
cancels out in the angular distributions,
{\em i.e.} the differential \xs s normalized to one.
We therefore choose for the processes (\ref{pe},\ref{ee})
to use normalized numbers of events as observables.

In contrast,
muon pair-production (\ref{mm})
does not have too large \xs s.
Moreover,
as this reaction proceeds solely via $s$-channel exchanges,
it is very sensitive to a possible $Z'$ tail.
It is therefore important to keep the information
about absolute rates.

For these reasons,
we use in our analysis
the following two observables:
\bea
\label{abs}
\pe\to\mu^+\mu^-~:
&&
A_i=\displaystyle{n_i}
\\
\label{rel}
\ee\to\ee~,~\pe\to\pe~:
&&
B_i=\displaystyle{n_i/N}
~,
\eea
where $N$ is the total number of events
and $n_i$ is the number of events within one angular bin,
labeled by the index $i$.
These numbers of events
are obtained by 
integrating the differential \xs s 
$d \sigma \over d \theta$
over the angular range
$[\theta_{-},\theta_{+}]$
which corresponds to bin $i$,
and by folding over 
%the detector angular resolution and 
the initial energy spectrum:
\bea
\nonumber
n_i & = &
{\cal L}
\displaystyle\int_0^1 \! dy_1 \, \Psi(y_1,\sqrt{s})
\displaystyle\int_{{E_{\min}\over y_1\sqrt{s}}}^1 \! dy_2 \, \Psi(y_2,\sqrt{s})
%\\\nonumber&&\hskip-0.5em
%\displaystyle\int_0^\pi \! d\theta_{-} \, \Phi(\theta_{-},\theta_{-}^0,\Delta\theta)
%\displaystyle\int_{\theta_{-}}^\pi \! d\theta_{+} \, \Phi(\theta_{+},\theta_{+}^0,\Delta\theta)
\\&&\hskip5em
\label{fold}
\displaystyle\int_{\theta_{-}^i}^{\theta_{+}^i} \! d\theta \, {d \sigma(y_1y_2s) \over d \theta}
~.
\eea
The integrated luminosity is given by $\cal L$.

\begin{comment}{
For the angular smearing $\Phi$
we use the Breit-Wigner distribution
\beq
\label{bw}
\Phi(\theta,\theta^0,\Delta\theta) = 
{ \Delta\theta/\pi \over (\theta - \theta^0)^2 + \Delta\theta^2 }
~,
\eeq
where $\theta^0$ is the {\em measured} edge of the angular bin,
$\theta$ is its {\em true} value
and $\Delta\theta$ is the angular resolution of the detector.
The Breit-Wigner error distribution is more conservative
than a gaussian
and is more convenient numerically.
}\end{comment}

At this stage we only consider the effect of initial state radiation
for the energy spectrum $\Psi$ of the initial electrons.
This means that
we ignore the effects of 
beamstrahlung 
and the beam energy spread
after bunch compression.
Even though the initial state radiation
should be the dominant contribution to the energy spread,
this is by no means sufficient.
However, 
the savings in calculation time are substantial
and this should provide a feeling for the importance 
of these effects.
To implement the initial state radiation
we use for $\Psi$ 
the Kuraev-Fadin spectrum~\cite{kf}.
\begin{comment}{
an inverted Weisz\"acker-Williams spectrum.
This spectrum is normalized to unity
by a box function
in the vicinity of $y=1$.
}\end{comment}

In the linear approximation
the error contours 
corresponding to 95\%\ confidence levels
are given by the quadratic form
in $v_{Z'}^2$ and $a_{Z'}^2$
\beq
\label{egg}
\left(
~v_{Z'}^2~
~a_{Z'}^2~ 
\right)
~~ W^{-1} ~~
\left(
\barr{c} 
v_{Z'}^2\\\\
a_{Z'}^2
\earr
\right)
~=~6
~,
\eeq
where the inverse covariance matrix $W^{-1}$ is given by
\beq
\label{cov}
W^{-1}_{ab} 
=
\sum_{\rm polarizations}~
\sum_{i=1}^{\rm bins}
{1\over\Delta O_i^2}
\left( {\partial O_i \over \partial \epsilon_a} \right)
\left( {\partial O_i \over \partial \epsilon_b} \right)
\eeq
and
\beq
\left\{
\begin{array}{l}
\epsilon_{a,b} = v_{Z'}^2,a_{Z'}^2
\\\\
O_i = A_i,B_i
~.
\end{array}
\right.
\eeq
For each observable (\ref{abs},\ref{rel})
the squared errors in the denominator of Eq.~(\ref{cov}) 
are given by the quadratic sum of the statistical and systematic errors
\bea
\label{relerr}
\hskip-1em
\Delta A_i^2
&=&
\displaystyle{n_i}
~+~
\sum_a 
\left( {\partial A_i \over \partial \eta_a} \Delta \eta_a \right)^2
\\
\label{abserr}
\hskip-1em
\Delta B_i^2
&=&
\displaystyle{n_i/N^2} \left( 1 - \displaystyle{n_i/N} \right)
~+~
\sum_a 
\left( {\partial B_i \over \partial \eta_a} \Delta \eta_a \right)^2
~,
\eea
where the first terms on the right-hand-side
are the statistical error.
The second terms include the systematic errors
on the bin edges $\theta^i_\pm$,
the polarizations $P_{1,2}$ 
and the integrated luminosity $\cal L$,
all added in quadrature.
The systematic errors
tend to cancel out 
for the ratios $B_i$ (\ref{rel}).
In particular
the error on the luminosity
cancels out exactly,
of course.

It is not a good approximation here,
to assume a linear dependence of the observables
on the parameters,
in this case 
$v_{Z'}^2$ and $a_{Z'}^2$.
But this simplication considerably accelerates the time needed 
for computing the error contours.
Moreover,
our prime interest here
is to evaluate the {\em relative} importance 
of the different systematic effects,
which should not be significantly influenced by this approximation.

\section{Results}

We concentrate our numerical analysis
on the first stage of the NLC,
with a \cm\ energy
\beq
\sqrt{s} = 500 \mbox{ GeV}
~.
\eeq

For the luminosities and their errors
we take
\beq
\begin{array}{l}
{\cal L}_{e^+e^-} = 50 \mbox{ fb}^{-1}
\\\\
{\cal L}_{e^-e^-} = 25 \mbox{ fb}^{-1}
\\\\
\displaystyle{\Delta{\cal L} \over {\cal L}} = 0.5\%
~.
\end{array}
\eeq
Since in the LEP1 experiments
the errors on the luminosity measurements
are approximately 0.3\%\ 
and are dominated by the theoretical error,
this choice is quite conservative for the NLC.
The lower luminosity
of the \ee\ mode
is due to the anti-pinching at the interaction point~\cite{jim}.

We do not consider polarized positrons,
but we assume for the electron beam polarizations 
the realistic future values
\beq
\begin{array}{l}
P_{e^-} = \pm 0.9
\\\\
\displaystyle{\Delta P_{e^-} \over P_{e^-}} = 1\%
~.
\end{array}
\eeq
For the \pe\ processes,
there are thus only two combinations of polarizations 
to sum over in Eq.~(\ref{cov}):
$e^+e^-_L$ and $e^+e^-_R$.
For M\o ller scattering,
there are three possibilities:
$e^-_Le^-_L$, $e^-_Re^-_R$ and $e^-_Le^-_R$.
We do not consider the third one,
since it is not very sensitive 
to the existence of a $Z'$~\cite{ccl}.

If the NLC and LEP detectors are comparable,
we may assume an angular coverage and resolution 
of
\beq
\begin{array}{l}
|\cos\theta| < 0.985
\\\\
\Delta\theta = 10 \mbox{ mrad}
~.
\end{array}
\eeq
We subdivide this angular range 
into 10 equal size bins 
in the cosine of the polar angle.
This number is only dictated by the savings in computation time.
A larger number of bins would only yield better results.
The formal limit of an infinite number of infinitesimal bins
is equivalent to a maximum likelihood fit 
and would yield the Cram\'er-Rao bound~\cite{cr}.

Finally,
to avoid the backgrounds from two-photon processes,
for instance,
we demand the total energy of the final state lepton pair
to be sufficiently close to the \cm\ energy
\beq
E_{\rm event} > \sqrt{s} - 10 \mbox{ GeV}
~.
\eeq

\begin{figure}[ht]
\hskip-2em
\input{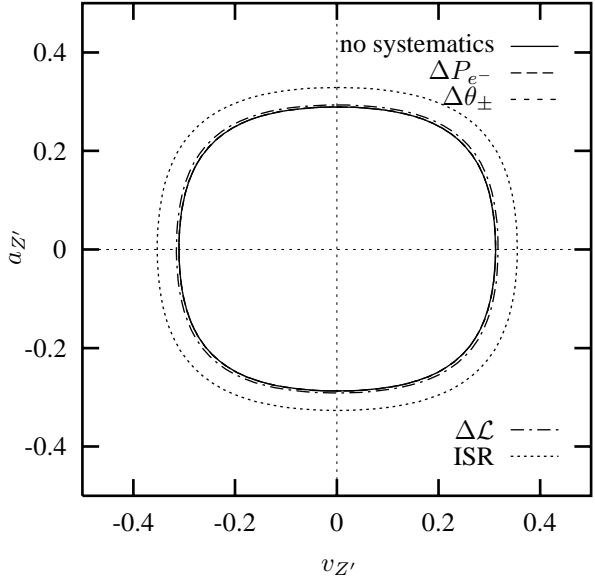}
\vskip-2ex
\caption{
Contours of observability at 95\%\ confidence
of the $Z'$ couplings to the charged leptons
for the reaction
$\pe\to\mu^+\mu^-$.
Results are individually shown 
for each of the four systematic effects
considered in the text,
and in the absence of any.
The curves corresponding to the inclusion
of the polarization error 
and the angular resolution
cannot be distinguished from the ideal case 
on this scale.
}
\label{fmm}
\end{figure}

\begin{figure}[ht]
\hskip-2em
\input{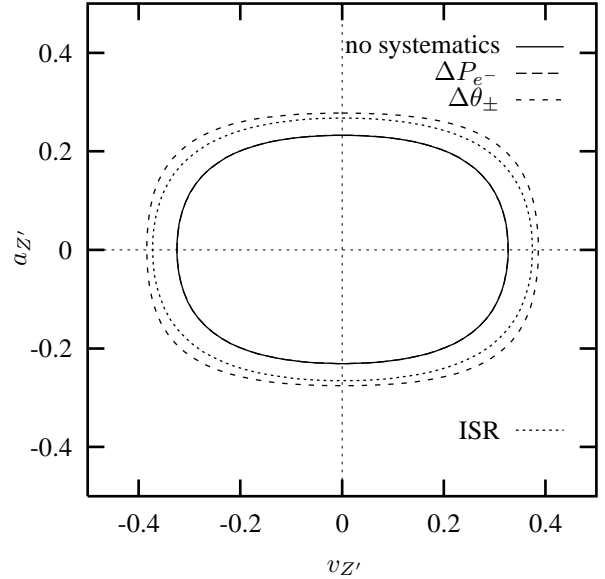}
\vskip-2ex
\caption{
Same as Fig.~\protect\ref{fmm},
for the reaction
$\pe\to\pe$.
The curve corresponding to the inclusion
of the polarization error 
cannot be distinguished from the ideal case 
on this scale.
}
\label{fpe}
\end{figure}

We plot for each reaction (\ref{mm}--\ref{ee})
in Figs~\ref{fmm}--\ref{fee}
the $(v_{Z'},a_{Z'})$ contours
defined by the quartic form (\ref{egg}).
In the absence of a signal
the parameter areas outside the contours
are excluded to better than 95\%\ confidence.
These results are obtained
assuming a $Z'$ mass of 2 TeV.
For other choices
the contours approximately scale according to Eq.~(\ref{scale}).

Even with our quite conservative assumptions 
for the systematic errors,
the resolving power of muon pair-production (\ref{mm}) 
remains dominated by the statistical errors.
The effect of initial state radiation
amounts to an increase by 15\% 
of the lower limit on the observable $Z'$ couplings.

Bhabha scattering (\ref{pe}) 
suffers a similar loss of resolving power
due to initial state radiation.
However,
the finite detector angular resolution 
is now the major source of degradation in accuracy.
This is due to the fact 
that this reaction is most sensitive to the presence of a $Z'$ 
in the forward scattering region,
where the effect of the angular error is largest.
The dependence on the polarization,
though,
is not a matter of concern.

M\o ller scattering (\ref{ee})
is mildly sensitive to all systematic effects.
In order of increasing importance
we have
the angular resolution,
the error on the polarization
and the smearing due to initial state radiation.
As for muon pair-production and Bhabha scattering,
the latter 
decreases the sensitivity by no more than 15\%.

\begin{figure}[ht]
\hskip-2em
\input{ee.pstex}
\vskip-2ex
\caption{
Same as Fig.~\protect\ref{fmm},
for the reaction
$\ee\to\ee$.
}
\label{fee}
\end{figure}

\begin{figure}[ht]
\hskip-2em
\input{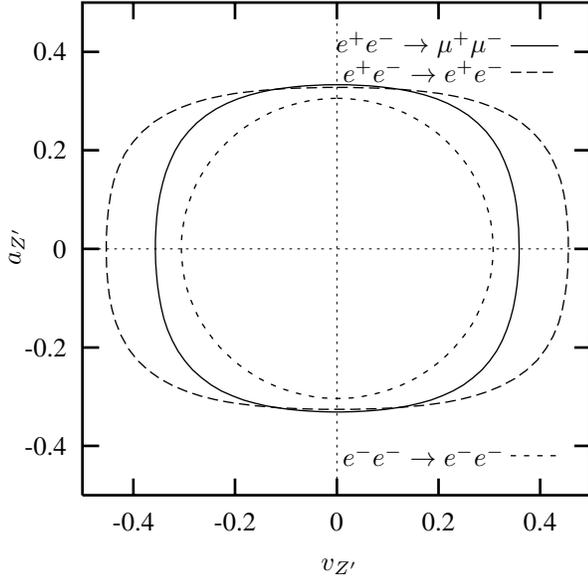}
\vskip-2ex
\caption{
Same as Fig.~\protect\ref{fmm},
for the three reactions (\protect\ref{mm}--\protect\ref{ee}).
All the systematic effects discussed in the text are included.
}
\label{fall}
\end{figure}

To gauge the discovery potential 
of the three reactions (\ref{mm}--\ref{ee})
we plot
their 95\%\ confidence discovery contours
together
in Fig.~\ref{fall}.
These contours include
the effects of all four systematic sources of accuracy loss
we have considered here.

\section{Conclusions}

Most current analysis of $Z'$ searches at the NLC
are performed for lepton or quark pair-production.
We have considered here the prospects of
Bhabha and M\o ller scattering
and have compared them with
muon pair-production.
The three reactions
turn out to
have a similar discovery potential.
M\o ller scattering is slightly more performant
than the two other processes.

All three reactions 
are relatively insensitive to systematic effects.
The most important degradation of the resolving power
is due to the finite angular resolution 
for Bhabha scattering,
and to the initial state radiation 
for M\o ller scattering and muon pair-production.
The lower discovery limits to be set on the leptonic $Z'$ couplings
worsens 
with respect to the case of an ideal beam and detector
by about 15\%, 41\% and 26\%
for muon pair-production, 
Bhabha scattering 
and M\o ller scattering
respectively.

o perform this study
we have used a linear approximation
and we have ignored the effects 
of the beam energy spread
and of beamstrahlung.
Our results are,
therefore,
correct only at the qualitative level.
It is unlikely,
though,
that the full calculation
will yield significant departures from these conclusions.

%%%%% References
%

\end{document}